\documentclass[usenatbib]{mn2e}
\usepackage{graphicx,natbib}
\usepackage{amsmath}

\newcommand{\msun}{M_{\odot}}\newcommand{\tmsun}{\mbox{$\msun$}}
\newcommand{\PN}{$\mathcal{PN}$}
\newcommand{\myvec}[1]{{\boldsymbol #1}}
\newcommand\simless{\mathbin{\lower 3pt\hbox
   {$\rlap{\raise 5pt\hbox{$\char'074$}}\mathchar"7218$}}} 
\newcommand\simgreat{\mathbin{\lower 3pt\hbox
   {$\rlap{\raise 5pt\hbox{$\char'076$}}\mathchar"7218$}}} 

\title[Tracing IMBHs in the Galactic Centre]
      {Tracing intermediate-mass black holes in the Galactic Centre}
      
\author[U. L\"ockmann and H. Baumgardt]
{
  U. L\"ockmann\thanks{E-mail: uloeck@astro.uni-bonn.de (UL); holger@astro.uni-bonn.de (HB)}
  and H. Baumgardt\footnotemark[1]\\ 
  Argelander Institute for Astronomy, University of Bonn, Auf dem H\"ugel 71, 53121 Bonn,
  Germany\\
}

\begin{document}

\date{Accepted 2007 November 8. Received 2007 November 8; in original form 2007 October 1}

\pagerange{323--330} \pubyear{2008} \volume{384}

\maketitle

\label{firstpage}

\begin{abstract}
We have developed a new method for post-Newtonian, high-precision integration of stellar systems
containing a super-massive black hole (SMBH),
splitting the forces on a particle between a dominant central force and perturbations.
We used this method to perform fully collisional $N$-body simulations of
inspiralling intermediate-mass black holes (IMBHs)
in the centre of the Milky Way.
We considered stellar cusps of different power-law indices
and analysed the effects of IMBHs of different masses, all starting from circular orbits at an
initial distance of 0.1\,pc.

Our simulations show how IMBHs deplete the central cusp of stars,
leaving behind a flatter cusp with slope consistent with what has recently been observed.
If an additional IMBH spirals into such a flat cusp, it can take 50\,Myr or longer to merge with the central SMBH,
thus allowing for direct observation in the near future.
The final merger of the two black holes involves gravitational wave radiation which may be observable with planned gravitational wave detectors.

Furthermore, our simulations reveal detailed properties of the hypervelocity stars (HVSs)
created, and how generations of HVSs can be used to trace IMBHs in the Galactic Centre.
We find that significant rotation of HVSs (which would be evidence for an IMBH) can only be expected
among very fast stars ($v > 1000$\,km\,s$^{-1}$).
Also, the probability of creating a hypervelocity binary star is found to be very small.

\end{abstract}


\section{Introduction}

In recent years, observations of the Galactic Centre revealed a number of interesting phenomena.
Analysis of the motions of stars within 1.2 arcsec of the Galactic Centre, for example,
showed that the Milky Way contains a super-massive black hole (SMBH) of mass $\approx 3.5 \times 10^6\,\msun$,
surrounded by an extremely dense stellar cluster \citep{sch03, ghe05}.

Since most massive galaxies are now believed to host such SMBHs \citep{mag98, getal00},
due to its proximity ($\approx$ 8\,kpc), the centre of the Milky Way provides a unique environment
to study the dynamics of SMBHs and the surrounding stellar systems.

While SMBHs have been directly observed as radio sources and compact massive objects,
the existence of black holes of masses $10^2\,\msun < M_{\rm BH} < 10^4\,\msun$,
so-called intermediate-mass black holes (IMBHs), is still a matter of debate
\citep[e.g. ][]{hpa04, mc04, pbhmm04}.

IMBHs could form in galaxies through runaway collisions of stars
in star clusters, giving rise to ultra-luminous X-ray sources
\citep{gfr04,pbhmm04,bhpm06}. They could later be brought into the centres of galaxies
through dynamical friction of the star clusters \citep{pbmmhe06}.
Some of the dynamical processes at the Galactic Centre may be well explained
by the presence of an IMBH, such as the existence and kinematics of young stars in the central 0.1\,pc \citep{hm03}.
Mergers of IMBHs with the central SMBH would be an important source
of gravitational waves, detectable with the next generation of gravitational wave detectors,
like e.g. the \emph{Laser Interferometer Space Antenna} \citep[\emph{LISA},][]{lisa}.

During the inspiral process of an IMBH, close encounters with stars from the central cluster eject some of them with velocities
large enough to leave the gravitational potential of the Milky Way,
 so-called hypervelocity stars (HVSs).
\citet{h88} was the first to show that the ejection of HVSs is a natural consequence of galaxies
hosting SMBHs.
Recently, several HVSs have been discovered in the Galactic halo \citep{b05, hi05, b06a, b06b, b07a, b07b}.
Except for one star which might have been ejected from the Large Magellanic Cloud
\citep{edel05}, the travel times of all HVSs are short enough 
that the stars could have been ejected from the Galactic Centre 
within their lifetimes, confirming Hills' predictions.
\citet{bgp06} and \citet{l06} have shown that HVSs may be ejected almost isotropically
by short bursts of inspiralling IMBHs, explaining the near-isotropic distribution of the so far detected HVSs \citep{b07a}.
However, \citet{shm07} found that the relatively small medium velocity of the (admittedly limited) sample of HVSs detected so far
favours other ejection mechanisms like the tidal breakup of stellar binaries \citep{h88, yt03, pha06}
or the scattering of stars by stellar-mass black holes \citep{mg00, oll06}.

In this work, we present new results on the inspiral of IMBHs, using
a novel method for studying the dynamics of Galactic Centre-like systems
where stars orbit a central SMBH on weakly perturbed Keplerian orbits.
In Section \ref{sec:bhint}, we describe the details of our integration method.
In Section \ref{sec:models}, we describe our set of N-body runs with IMBHs in the Galactic Centre.
Our results are given in Section \ref{sec:results} and discussed in Section \ref{sec:discussion}.

\section{Integrator for SMBH-dominated systems}
\label{sec:bhint}

\subsection{Requirements}

    Stellar systems around SMBHs are a unique environment:
    Dynamically, they behave like the solar system in that stars move along weakly perturbed Keplerian orbits
    around a massive central body.
    Integration of planetary systems is usually done by symplectic integrators,
    which yield a very good energy conservation for nearly circular orbits
    but fail for eccentric orbits with a wide range of semimajor axes.

    The structure of the stellar system around a massive black hole is also similar to a star cluster,
    since it comprises a large number of stars of similar mass with wide ranges of eccentricities and central distances.
    Good progress has been made in the development of high precision integrators
    for the dynamical simulation of star clusters, usually based on the Hermite algorithm \citep{ma92}.

    A drawback underlying Hermite integrators is their inability to differentiate massive objects
    dominating the system's dynamics from relatively small masses 
    causing only small perturbations to the particles' orbits. 

    Although integrators have been developed in both the areas of planetary systems and stellar clusters,
    a new approach is therefore required for the Galactic Centre.

    We have developed an integrator \textsc{bhint} specialized for dynamical processes
    in the vicinity of a SMBH,
    making use of the information that the black hole dominates the motion of the stars.
    For this new mechanism, we retain the Hermite scheme as a basis,
    providing very fast calculations on the {\it GRAPE} special-purpose hardware \citep{mfkn03}
    which is essential for the large values of $N$ we use.

\subsection{Basic integration algorithm}

The basic idea for our new method is to split the force calculation between the dominating central force (exerted by the SMBH)
and the perturbing forces (due to the cluster stars). It is comparable to the so-called \emph{mixed variable symplectic} methods
\citep[\emph{MVS,}][]{wh91}, as it makes use of Kepler's equation to integrate along the orbit. However, our method is not symplectic, 
as it is based on the Hermite scheme to allow for use of the {\it GRAPE} special-purpose hardware.
Furthermore, we assume the dominating SMBH to rest at the centre and hence do not need Jacobi coordinates but calculate the orbital
motion in Cartesian coordinates.

A different advancement of the idea of \emph{MVS} was also made recently by \citet{fifm07} for simulation of a star cluster and its parent galaxy,
separating forces between star cluster particles and galaxy particles.

The underlying Hamiltonian for a Newtonian problem of $N$ particles with masses $m_0..m_{N-1}$,
positions $\myvec{r}_0..\myvec{r}_{N-1}$ and velocities $\myvec{v}_0..\myvec{v}_{N-1}$ is
\begin{equation}
  H = \sum_{i=0}^{N-1}{\frac{m_i\myvec{v}_i^2}{2}} + \sum_{i<j}{\frac{Gm_im_j}{r_{ij}}}
\end{equation}
where $\myvec{r}_{ij} := \myvec{r}_{j} - \myvec{r}_{i}$, $r_{ij} := |\myvec{r}_{ij}|$,
and $G$ is the gravitational constant.
As the central object is much more massive than the orbiting particles, we will assume
it does not move with respect to the system's centre of mass ($\myvec{r}_0 \equiv 0$).\footnote{
For our models, the expected motion of the SMBH relative to the system's centre of mass is of the order of $3\times 10^{-4}$\,pc.
Only few of the stars have a pericentre distance within that range, and most of these will move around with the SMBH,
as their orbital period is much shorter than the expected SMBH motion time-scale \citep[cf.][]{chl02}.
Hence, the motion of the SMBH can be neglected.}
In this case, the total force acting on a particle can be split into the unperturbed motion
around the central object, and the perturbing force, as
\begin{equation}
  \ddot{\myvec{r}}_i = \underbrace{-Gm_0\frac{\myvec{r}_i}{r_i^3}}_{\ddot{\myvec{r}}_{i,K}}
  \underbrace{-\sum_{\genfrac{}{}{0pt}{}{j=1}{j\not=i}}^{N-1}{Gm_j\frac{\myvec{r}_{ij}}{r_{ij}^3}}}_{\ddot{\myvec{r}}_{i,P}}
  \label{eq:forcesum}
\end{equation}
for $i=1\dots N-1$.

The new position of a particle after a time-step $\Delta t$ can be expressed as
\begin{eqnarray}
  \myvec{r}' 
  =& \myvec{r}
  + \dot{\myvec{r}} \Delta t
  &+ \frac{1}{2!} \ddot{\myvec{r}}_K \Delta t^2
  + \frac{1}{3!} \myvec{r}^{(3)}_K \Delta t^3
  + \dots \nonumber \\
  &&
  + \frac{1}{2!} \ddot{\myvec{r}}_P \Delta t^2
  + \frac{1}{3!} \myvec{r}^{(3)}_P \Delta t^3
  + \dots\label{eq_algo}
\end{eqnarray}

The first line of equation (\ref{eq_algo}) is solved directly, assuming an unperturbed motion
along a Keplerian orbit and making use of Kepler's equation.
The second line is calculated using the Hermite integration scheme
as described in \citet{ma92}.
To adapt this scheme to our integrator, the predictor step for the particles to be moved
needs to be changed to
\begin{eqnarray}
  \myvec{r}'_{\rm pred} &=& \myvec{r}_{1,K}
  + \frac{1}{2} \myvec{a}_P \Delta t^2
  + \frac{1}{6} \dot{\myvec{a}}_P \Delta t^3 \\
  \myvec{v}'_{\rm pred} &=& \myvec{v}_{1,K}
  + \frac{1}{2} \myvec{a}_P \Delta t
  + \frac{1}{6} \dot{\myvec{a}}_P \Delta t^2
\end{eqnarray}
where $\myvec{r}_{1,K}$ and $\myvec{v}_{1,K}$ are the results of Kepler integration.

To determine the size of a particle's next time-step, the following criteria are used:
\begin{enumerate}
\item The particle's central motion.
  The accuracy of the Kepler orbit integration does not depend on the step size,
  however the faster change of the particle's velocity
  at perihelion requires a smaller time-step as it causes faster changes in perturbing forces.
  Hence we introduce a scale-independent upper time-step limit as
  \begin{equation}
    \Delta t_{\rm kep} = \frac{2 \pi}{N_{\rm step}}\sqrt{\frac{r^3}{GM_{\rm SMBH}}}
  \end{equation}
  where $r$ is the particle's central distance, $M_{\rm SMBH}$ is the SMBH mass and
  $N_{\rm step}$ a constant which we set equal to 50.
  For a circular orbit, $N_{\rm step}$ gives a lower bound to the number of steps per orbit
  (more steps are needed at the pericentre of eccentric orbits).
\item The relative change of perturbing forces.
  Here, we apply the criterion introduced by \citet{aar85},
  \begin{equation}
    \Delta t_{\rm pert}
     = \sqrt{\eta \frac{|\myvec{a}||\ddot{\myvec{a}}|+|\dot{\myvec{a}}|^2}
     {|\dot{\myvec{a}}||\myvec{a}^{(3)}|+|\ddot{\myvec{a}}|^2}},
  \end{equation}
  and use $\eta = 0.5$. \citet{aar85} suggests a value of $\eta = 0.02$; however,
  in our case changes of perturbations are usually of rather little effect on a particle's trajectory,
  as the orbital motion around the SMBH is dominating.
  Thus to retain a certain level of energy conservation, the time-steps can be much larger compared
  to a standard Hermite scheme, as the dominating orbital energy is conserved in any case by our method.
\item To prevent jumps in energy by not handling close encounters correctly, we need to consider
  changes in the particle's vicinity (i.e.\ approaches and close encounters).
  We assign a neighbour sphere to each particle and check for fast approaches and close encounters
  \emph{before} we perform a step. Thus, we prevent sudden changes of perturbing force derivatives.
  A particle's close encounter step size $\Delta t_{\rm enc}$ is set to the maximum of all step sizes,
  during which the particle will not approach any other particle by more than a factor of $\sim 3$.
  \label{item:appcheck}
\end{enumerate}

The definitive time-step of a particle $i$ is defined as the minimum of $\Delta t_{\rm kep}$, $\Delta t_{\rm pert}$,
and $\Delta t_{\rm enc}$, reduced to the next smaller integer power of two
\begin{equation}
  \Delta t_i = 2^{\lfloor\log_2{\operatorname{min}\left(\Delta t_{\rm kep,i}, \Delta t_{\rm pert,i}, \Delta t_{\rm enc,i} \right)}\rfloor}.
\end{equation}

In our simulations, all particles are treated as point masses.
We do not consider tidal disruption during our calculations, but track close encounters between two particles.

Our simulations show that binary formation in a SMBH-dominated cusp can be neglected,
therefore we do not use any regularization methods.

\subsection{Comparison with other integration methods}

High-precision symplectic integrators are available for different kinds of problems,
such as the time-symmetric adaptive time-step mechanism introduced by \citet{mt99},
which was advanced by \citet{lwt05} for use in a three-body problem
to study the effects of an inspiralling IMBH in the Galactic Centre on a single star,
or the individual time-step multiple variable symplectic integrator by \citet{st94},
which is designed for planetary systems.
However, symplectic integrators fail when dealing with Galactic Centre-like systems,
as these comprise a wide range of semimajor axes and eccentricities,
and no symplectic integrators with individual \emph{and} adaptive time-steps have been found as yet.

Therefore, we can only compare our new method to an integration scheme for large stellar clusters,
among which the Hermite method \citep{ma92} is the most widely used method for star cluster integration.

\begin{figure}
  \begin{center}
    \includegraphics[width=8.3cm]{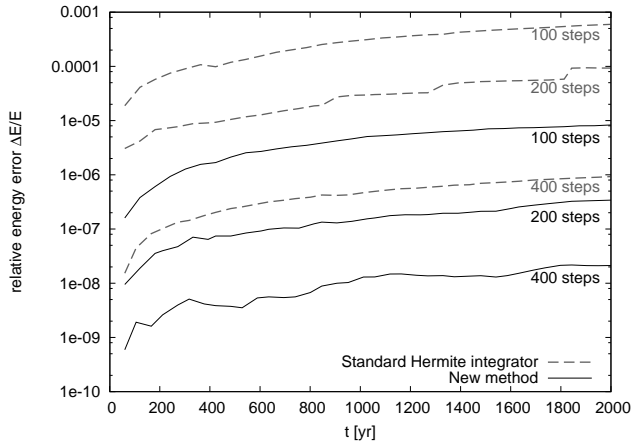}
  \end{center}
  \caption{Comparison of our new integrator method \textsc{bhint} (solid lines) with the standard Hermite scheme (dotted lines)
    for a cluster of 1000 stars with masses $ 3 - 10\,\msun$ on orbits with a range of eccentricities
    and semimajor axes between 0.05 and 0.1\,pc around a SMBH of mass $3.5 \times 10^6\,\msun$.
    For a given average number of steps per orbit, \textsc{bhint} is almost a factor of 100 more accurate than the standard Hermite method.
    \label{fig:int-comp}}
\end{figure}

Fig.\ \ref{fig:int-comp} compares the performance of our new method \textsc{bhint} with a standard Hermite scheme.
It shows the relative energy error for integration of a cluster of 1000 stars
with masses $3 - 10\,\msun$ on orbits with a range of eccentricities
and semimajor axes between 0.05 and 0.1\,pc around a SMBH of mass $3.5 \times 10^6\,\msun$.
For a given average number of $(100, 200, 400)$ steps per orbit, one can see that our method is almost a factor of 100
more accurate than the standard Hermite method,
as it does not accumulate an error in the orbital motion around the SMBH.

The secular energy error of the Hermite scheme has been investigated by \citet{kym98},
who also introduced a multiple-evaluation Hermite scheme, P(EC)$^n$, to overcome this drawback
for low-eccentricity systems with rare close encounters (such as planetary systems).
Our new method obviates the energy error of the orbital motion (which is calculated exactly by solving Kepler's equation),
but only retains the energy error due to interactions between the orbiting particles,
which in systems dominated by a central body is much lower than the total energy.

\subsection{Post-Newtonian extension}
As an IMBH spirals into the Galactic Centre due to dynamical friction,
it can eventually reach a highly eccentric orbit with a relatively small semimajor axis,
a regime where relativistic effects become important.
For example, according to equation (5.10) of \citet{p64}, a $3 \times 10^3\,\msun$ IMBH orbiting a $3.5 \times 10^6\,\msun$ SMBH
on a $10^{-3}$\,pc orbit with eccentricity 0.97
will merge with the SMBH within $\sim 1$\,Myr due to gravitational radiation.
 The actual value is even smaller, as the relativistic pericentre shift
prolongates the pericentre passage.

To account for relativistic effects, we extend our integration
method by post-Newtonian (\PN) correction terms.
This changes the calculated acceleration to
\begin{equation}
  \myvec{a} = \underbrace{\myvec{a}_0}_{\mathrm{Newtonian}}
  + \underbrace{\underbrace{c^{-2}\myvec{a}_2}_{1\mathcal{PN}}
    + \underbrace{c^{-4}\myvec{a}_4}_{2\mathcal{PN}}}_{\mathrm{pericentre\ shift}}
  + \underbrace{\underbrace{c^{-5}\myvec{a}_5}_{2.5\mathcal{PN}}}_{\mathrm{GW}}
  + \mathcal{O}(c^{-6}).
\end{equation}
The actual formulae for the \PN\ correction terms are taken from \citet[equation (168)]{bla06},
simplified to our case of a SMBH at rest.
To use these in our Hermite scheme for perturbing forces, we also calculated the force derivatives.
\citet{kas06} were the first to include these post-Newtonian corrections into a standard Hermite integration method.

While the 2.5\PN\ term corresponds to the emission of gravitational waves and thus does not conserve energy,
the 1\PN\ and 2\PN\ terms are responsible only for pericentre shift and do not change the semimajor axis
and eccentricity. However, even these two terms lead to a large periodic change in the Newtonian energy.

For these reasons, we track the integrator's energy conservation
similar to  \citet{aar07} by evaluating the path integral
\begin{equation}
  \Delta E = \int{m\myvec{v}\myvec{a}_{\mathrm{GW}}\,dt}.
\end{equation}

To account for relativistic effects, we replace the Keplerian force $\ddot{\myvec{r}}_{i,K}$ in equation (\ref{eq:forcesum})
by the relativistic Keplerian force
\begin{equation}
  \ddot{\myvec{r}}_{i,K_{\mathcal{PN}}} = -Gm_0\frac{\myvec{r}_i}{r_i^3}
  + c^{-2}\myvec{a}_2
  + c^{-4}\myvec{a}_4
  + c^{-5}\myvec{a}_5
  + \mathcal{O}(c^{-6}).
  \label{eq:pnforce}
\end{equation}
Since no analytical solution for the post-Newtonian two-body problem is known, we integrate this equation
using a standard Hermite scheme. In contrast to the Kepler equation, this is not exact, and especially since
the post-Newtonian forces change rather quickly, we replace every Kepler step by 256 steps of integrating
equation (\ref{eq:pnforce}). Only after performing these 256 steps, the perturbing forces are applied to a particle.

As the integration of the post-Newtonian terms is time-consuming and we are mainly interested in the IMBH inspiral,
we will only apply it to the IMBH, and only when its estimated merging time falls below 100\,Myr.
It is also possible to apply this integration to all particles below a certain threshold,
trading computational speed for higher accuracy of stars on close eccentric orbits.

\section{Description of the runs}
\label{sec:models}

All runs were performed with the code described above
on the GRAPE6 computers \citep{mfkn03} of Bonn University.
Our runs contained three different components: A central SMBH, an IMBH,
and a cluster of stars [varying in size and stellar mass,
but following the stellar mass estimated by \citet{gs03} within the inner 0.4\,pc].
In all simulations, the SMBH was at rest at the origin 
and had a mass of $M_{\rm SMBH} = 3 \times\,10^6\msun$, 
similar to the mass of the SMBH at the Galactic Centre 
\citep{sch03, ghe05}.
The mass of the IMBH was varied between $(1-3) \times 10^3\,\msun$ in the different runs.
All IMBHs moved initially in circular orbits at a distance of 0.1\,pc from the SMBH.
Details are shown in Table \ref{tab:runs}.

\begin{table}
  \begin{center}
    \begin{tabular}{|ccccc|}
    \hline
      Run & $M_{\mathrm{IMBH}}$ & $M_*$
        & $N$ & $\alpha$ \\
      \hline
      $A$ & $10^3$          & 30        &  19,470 & $1.75$\\
      $B$ & $3\times 10^3$   & 30        &  15,779 &  \\
      $C$ & $10^3$          & 30        &  13,549 &  \\
      $D$ & $3 \times 10^3 $ & $2.5-8.5$ &  58,483 & $1.4$\\
      \hline
    \end{tabular}
    \caption{Parameters used in N-body runs. Masses are given in \tmsun, $N$ is the number of particles.
      The SMBH mass is $3 \times 10^6 \msun$ in all runs.
      Runs $B$ and $C$ take the results of runs $A$ and $B$, respectively, as input, plus a follow-up IMBH.
      They thus start with an $\alpha = 1.75$ density profile at large distances and a slightly (run $B$)
      or strongly (run $C$) depleted cusp at smaller distances.
      \label{tab:runs}}
  \end{center}
\end{table}

The stellar cluster was set up such as to be initially non-rotating and fulfill an isotropic velocity distribution,
\begin{equation}
  \sum{v_r^2}=\sum{v_\theta^2}=\sum{v_\varphi^2}=\frac{1}{3}\sum{v^2}
\end {equation}
\citep[e.g.,][equation (4-53b)]{bt87},
which translates into a uniform distribution of $e^2$ and a mean central distance $\overline{r} = 5/4\ a$
(where $e$ and $a$ are a star's eccentricity and semimajor axis, respectively).
The initial density profile of the stellar cluster was a power law with different exponents for the different runs
(see Table \ref{tab:runs}).

\section{Results}
\label{sec:results}

\subsection{Predicted IMBH inspiral}
\label{sec:results:prediction}

Due to dynamical friction, an IMBH sinks into the centre of a stellar cusp.
The frictional drag on the IMBHs can be estimated by [see \citet{bt87}, equation (7-18)]:
\begin{equation}
 \frac{{\rm d}\myvec{v}}{{\rm d}t} = -\frac{4 \pi \ln{\Lambda} G^2 \rho(r) M_{\rm IMBH}}{v^3}
\left[ {\rm erf}(X) - \frac{2 X}{\sqrt{\pi}} e^{-X^2} \right] \myvec{v}
 \label{dynf}
\end{equation}
where $\rho(r)$ is the background density of stars,
$\ln{\Lambda}$ the Coulomb logarithm
and $X=\myvec{v}/(\sqrt{2}\sigma)$
is the ratio between the velocity of the IMBH and the (1D) stellar
velocity dispersion $\sigma$, which can be inferred from the model data.
For a stellar density profile $\propto r^{-\alpha}$ with $1.4 \leq \alpha \leq 1.75$, one gets $X \approx 1.2$.

For a power-law density profile $\rho(r)=\rho_0r^{-\alpha}$,
assuming the IMBH moves on a circular orbit, and introducing the constant
\begin{equation}
  A = 8\pi\ln\Lambda\frac{\sqrt{G}\rho_0M_{\mathrm{IMBH}}}{M_{\mathrm{SMBH}}^{3/2}}
  \left[\mathrm{erf}\left(X\right)-\frac{2X}{\sqrt{\pi}}e^{-X^2}\right]
  \nonumber,
\end{equation}
equation (\ref{dynf}) can be rewritten as
\begin{equation}
 F = M_{\mathrm{IMBH}} \left|\frac{{\rm d}\myvec{v}}{{\rm d}t}\right|
   = -\frac{v_c}{2} M_{\mathrm{IMBH}} A r^{3/2-\alpha}
    \label{eq:inspiral_force}.
\end{equation}

Since the rate of angular momentum change is equal to ${\rm d}L/{\rm d}t=F r/M_{\rm IMBH}$
and since the angular momentum itself is given by
$L = r v_c = \sqrt{G M_{\rm SMBH} r}$, we have
\begin{equation}
  \frac{F r}{M_{\rm IMBH}} = \frac{{\rm d}L}{{\rm d}t}
  = \frac{{\rm d}}{{\rm d}t} \sqrt{G M_{\rm SMBH} r}
  = \frac{v_c}{2} \frac{{\rm d}r}{{\rm d}t},
\end{equation}
and inserting equation (\ref{eq:inspiral}) leads to
\begin{equation}
  \frac{{\rm d}r}{{\rm d}t} r^{\alpha-5/2} = -A.
\end{equation}

Solving this equation with initial condition $r(0)=r_0$ gives the radius reached at time t:
\begin{equation}
  r(t)=\left(r_0^{\alpha-3/2}
    -A
    \left(\alpha - 3/2\right)t\right)^{\frac{1}{\alpha-3/2}}
    \label{eq:inspiral}.
\end{equation}

From this it can be seen that for cusps steeper than $\alpha = 1.5$, the inspiral process terminates in finite time,
whereas in shallower cusps the IMBH will never reach $r = 0$.

\subsection{IMBH inspiral in a Bahcall-Wolf cusp around a SMBH}
Theoretical arguments and N-body simulations have shown that a stellar system around a SMBH
evolves into a cusp with an $\alpha = 1.75$ power-law density distribution
\citep{bw76,bme04a,bme04b,pms04}.

\begin{figure}
  \begin{center}
    \includegraphics[width=8.3cm]{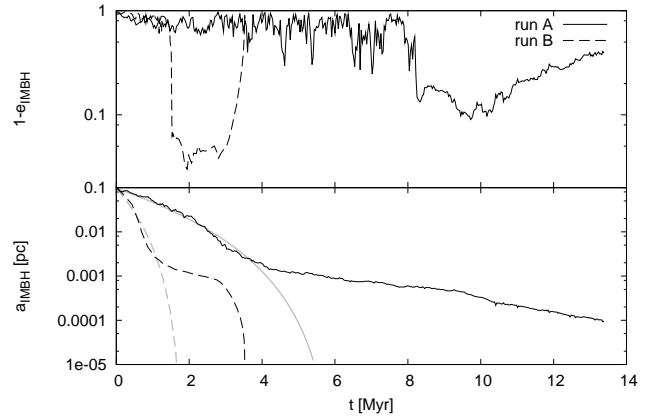}
  \end{center}
  \caption{Evolution of the inspiralling IMBHs' eccentricities $e_{\rm IMBH}$ and semimajor axes $a_{\rm IMBH}$.
    The solid and dashed lines depict the results of runs $A$ ($M_{\rm IMBH} = 10^3\,\msun$) and $B$ ($M_{\rm IMBH} = 3 \times 10^3\,\msun$), respectively,
    while the theoretical predictions are plotted in grey.
    Once the eccentricity increases, GW emission becomes important at the pericentre,
    leading to circularization of the orbit.
    For run $A$, we terminate the calculation after 12\,Myr.
    From the orbital parameters at this point, we estimate that the 
    inspiral time due to gravitational wave radiation is of the order of a few Myr.
    Run $B$ terminates as relativistic effects lead to SMBH-IMBH coalescence.
    \label{fig:a-imbh1000}}
\end{figure}

Following these theoretical predictions, we started our calculations
with a $10^3\,\msun$ IMBH on a circular orbit
around a $3\times10^6\,\msun$ SMBH with semimajor axis $a_{\mathrm{IMBH}} = 0.1$\,pc (run $A$).
Fig.\ \ref{fig:a-imbh1000} shows that the inspiral process follows the theoretical description
(equation (\ref{eq:inspiral})) very well.
We find a best fit for all our runs if we assume a Coulomb logarithm of $\ln{\Lambda} = 7.5$, which is in good agreement with \citet{mme07}.
The actual inspiral deviates from the prediction once the IMBH has started to deplete the inner cusp and the
density profile no longer follows an $r^{-1.75}$ law.
These results agree very well with those obtained by \citet{bgp06}.

From this point, the IMBH's central distance decreases only slowly, as two-body encounters are rare.
Gravitational wave (GW) emission at the pericentre becomes important after 8\,Myr,
when the IMBH acquires a highly eccentric orbit due to interactions with passing stars \citep{ha06b,mme07}.
This leads to further decline of the semimajor axis, and later on to a circularization of the orbit.
The inspiral time due to GW radiation is about 1\,Myr \citep[using equation\ (5.14) from][]{p64},
but orbit circularization delays this process.

After the IMBH in run $A$ has started to deplete the inner cusp and reached a central distance of $5 \times 10^{-4}$\,pc,
we start a subsequent run ($B$), assuming that a follow-up $3 \times 10^3\,\msun$ IMBH starts from a distance of 0.1\,pc.
Due to the higher mass, the inspiral process is much faster (Fig.\ \ref{fig:a-imbh1000}).
After 1\,Myr, the IMBH reaches the innermost region, where the density is decreased and the inspiral process is slowed.
After 1.5\,Myr, the IMBH acquires a highly eccentric orbit (up to $e=0.97$), and GW emission becomes important, leading to further decline of the semimajor axis.
After 3\,Myr, GW emission has taken over to rapidly decrease the semimajor axis and circularize the orbit to almost $e=0$.
Eventually, the pericentre distance falls below three SMBH Schwarzschild radii, and the IMBH is considered to be swallowed.

\subsection{IMBH inspiral in a flattened cusp}
Recent observations suggest that at least the density profile of the brightest stars in the Galactic Centre follows a more flattened cusp
than the $\alpha = 1.75$ profile assumed above \citep[e.g.,][]{gs03,sea+07}.
This may be the result of mass segregation and the preferential depletion of main-sequence stars
due to more massive stellar mass black holes \citep{bme04b,fak06}.
However, as can be seen in \citet{bgp06}, this can also be the result of a recent IMBH inspiral
(and thus apply to the whole stellar population in that region).

\citet{bgp06} estimated that replenishment of the cusp after an IMBH inspiral 
will take at least 100\,Myr, while \citet{pbmmhe06} calculated
an IMBH inspiral rate of 1 per 10\,Myr into the Galactic Centre.
This would explain the observed flattening of the stellar cusp in the centre of the Milky Way,
as a follow-up IMBH spirals in and scatters away stars before the cusp has been replenished.

In another subsequent model, we study the changes to an IMBH inspiral in such a flattened cusp.
A second follow-up IMBH (run $C$, $M_{\rm IMBH}=10^3\,\msun$) finds the density profile in the inner 0.01\,pc
to be flattened to $\propto r^{-1.2}$ by the preceding two IMBHs, causing the inspiral process to almost stop at a distance of $10^{-3}$\,pc,
where the effects of general relativity are still not sufficient to drive a merger with the SMBH within 1\,Gyr.

\begin{figure}
  \begin{center}
    \includegraphics[width=8.3cm]{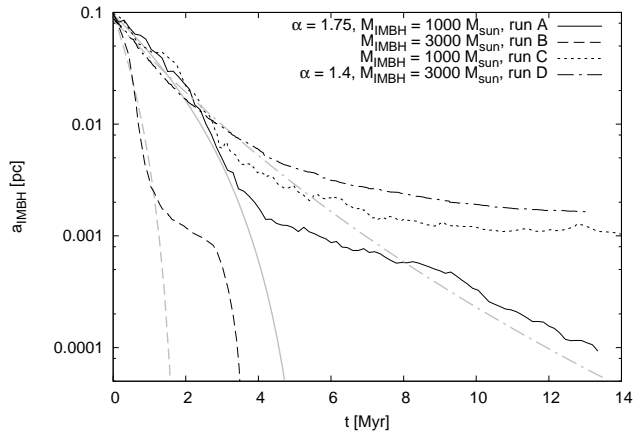}
  \end{center}
  \caption{
    Evolution of the inspiralling IMBHs' semimajor axes $a_{\rm IMBH}$.
    The solid, dashed, and dotted lines represent subsequent inspiralling IMBHs of mass
    $10^3\,\msun$ (run $A$),
    $3 \times 10^3\,\msun$ (run $B$),
    and $10^3\,\msun$ (run $C$), respectively.
    Due to the progressive depletion of stars in the inner cusp, the inspiral of the subsequent
    IMBHs terminate at larger radii (compare curves for run $A$ and $C$ with otherwise identical parameters).
    The dash-dotted line depicts the $3 \times 10^3\,\msun$ IMBH in an $\alpha=1.4$ cusp (run $D$).
    The actual inspirals deviate from the theoretical predictions plotted in grey once the
    inner cusp becomes depleted of stars. \label{fig:a-imbh-inspiral}}
\end{figure}

To reproduce the situation of a density profile
as found for the centre of the Milky Way by \citet{gs03},
we also analysed the inspiral of a $3 \times 10^3\,\msun$ IMBH in an $r^{-1.4}$ power-law cusp (run $D$).
To follow the dynamical processes more accurately, this time we used stellar masses between 2.5 and 8.5\,\tmsun.
According to equation (\ref{eq:inspiral}), for a power law with index $\alpha < 1.5$, the Newtonian decrease of semimajor axis
will never terminate.

Fig.\ \ref{fig:a-imbh-inspiral} shows the evolution of the IMBHs' semimajor axes in all four runs.
As we have also seen for the previous runs, our actual inspiral curve is in good agreement with the theoretical prediction for the flattened cusps.
The inspiral process is slowed as the IMBH clears out the central region.
Due to the lower number of stars in the centre of this shallower cusp, the decrease of semimajor axis terminates at a larger central distance.

\subsection{Detection of IMBH inspirals}

\begin{figure}
  \begin{center}
    \includegraphics[width=8.3cm]{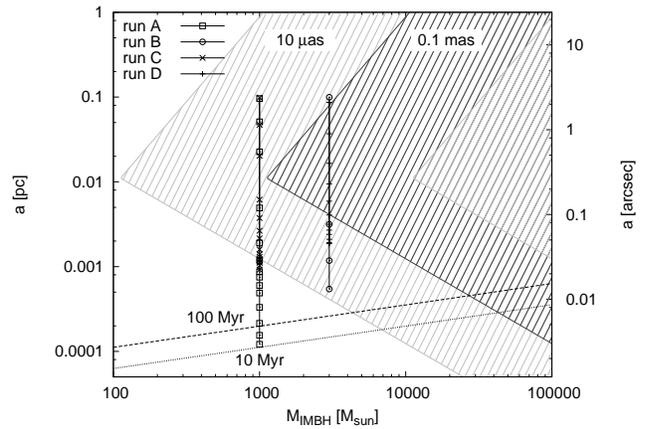}
  \end{center}
  \caption{Detection limits of an astrometric wobble of the radio image of Sgr A* as a function of IMBH mass,
  assuming a circular orbit, a SMBH mass of $3\times 10^6\,\msun$, and a distance to the Galactic Centre of 8.5\,kpc (adapted from \citet{hm03}).
  The shaded regions show the detection limits in semimajor axis $a$ and IMBH mass $M_{\rm IMBH}$ 
  for astrometric resolutions of 10\,$\rm \mu$as, 0.1\,mas, and 1\,mas, respectively, assuming a monitoring period of 10 yr.
  The dashed and dotted lines indicate coalescence due to gravitational wave radiation in 10$^7$ and 10$^8$\,yr, respectively,
  assuming a circular orbit.
  The motion of the IMBHs from runs $A$ to $D$
  is represented by solid lines, symbols depict steps of 1\,Myr.
  One can see that the $3\times 10^3\,\msun$ IMBH in run $D$ spends 5\,Myr in a regime where it would be detectable with
  an astrometric resolution of 0.1\,mas, and probably a few 10\,Myr in the 10\,$\rm \mu$as regime.
  \label{fig:astrometric}}
\end{figure}

\citet{hm03} have shown that an IMBH of $3 \times 10^3\,\msun$ on a circular orbit around Sgr A*
at a distance between $4 \times 10^{-3}$ and $8 \times 10^{-2}$\,pc may be observed by detecting the proper motion of Sgr A*
if the astrometric resolution reaches 0.1\,mas or better.
Fig.\ \ref{fig:astrometric} shows the amount of time the IMBHs in our simulations spend within certain detectability limits.
One can see that the IMBH in the flattened cusp (run $D$) 
spends several Myr in this regime,
yielding good chance of direct observation in the near future
if the inspiral rate of IMBHs is large enough.

At an orbital frequency of 10$^{-4}$\,Hz, the black hole binary enters the \emph{LISA} detection band.
Using the values from run $B$, this corresponds to a semimajor axis of $a=3.25 \times 10^{-6}$\,pc,
and about 2.5\,yr until coalescence.
Assuming a distance of $R=1$\,Gpc from the source, the amplitude of the GW signal for the circularized binary is given by \citep[equation (3.5) and (3.12)]{db79}
\begin{equation}
  h = \sqrt{\frac{128}{5}}\frac{G^2M_{\rm SMBH}M_{\rm IMBH}}{c^4 a R} \approx 3 \times 10^{-20}
\end{equation}
which is a factor of 3 above the noise limit \citep{skn01}.
Assuming a density of SMBHs with masses $10^6 \msun < M_{\rm SMBH} < 10^7 \msun$
of order ${\rm d}N/{\rm d}\log M_{\rm SMBH} = 3\times 10^6$\,Gpc$^{-3}$ \citep{ar02,gd07}
and thus a total number of $N_{\rm SMBH} \approx 1.3 \times 10^7$ within a distance of 1 Gpc,
as well as an IMBH inspiral rate of 1 per 25\,Myr, we get a 73 per cent chance to find a signal at a given point in time.
For an observation period of three years, this value rises to 94 per cent, yielding a good chance to observe IMBH inspirals with \emph{LISA}.
Furthermore, SMBHs in other mass ranges will contribute to the event rate.

\subsection{HVS ejection}
During the inspiral process, a number of HVSs stars are ejected from the system
by close encounters with the IMBH.
\citet{bgp06} derived relatively short HVS burst intervals of only a few Myr for IMBH inspirals in an $\alpha = 1.75$ power-law cusp.
However, if the IMBH inspiral terminates at a larger distance (as we have seen above for a shallower cusp as found in our Galaxy),
the ejection rate of HVSs will be lower, and HVSs will be created over a longer time interval.
For example, the HVS ejection rate in run $C$ reaches 4-7\,Myr$^{-1}$ during the first 5\,Myr of HVS ejection, but then retains an average rate of
3\,Myr$^{-1}$ throughout the rest of the simulation.
This could explain the absence of bursts in the observed distribution of Galactic HVSs \citep{b07b}.

To compare the results of our simulations with the HVSs found so far,
we followed the trajectories of all ejected stars with velocities $v > 275$\,km\,s$^{-1}$ at distances between 10 and 120\,kpc from the Galactic Centre.
For this range, \citet{b06b, b07a, b07b} made a complete survey of B-type stars across 12 per cent of the sky.
We used a Runge-Kutta integrator using the Galactic potential from \citet{p90},
adding the broken power-law density profile determined for the Galactic Centre by \citet{gs03} in the inner 40\,pc,
where the latter exceeds the density of the \citet{p90} model.

\begin{figure}
  \begin{center}
    \includegraphics[width=8.3cm]{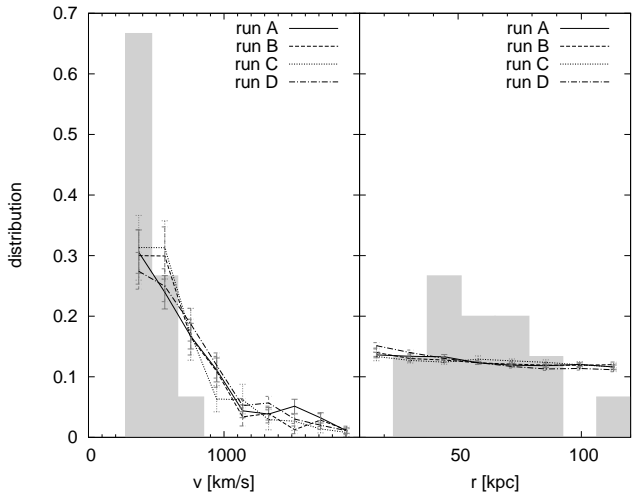}
  \end{center}
  \caption{Distribution of galactocentric velocities and distances of HVSs created in our simulations.
    We consider all ejected stars in the distance range 10\,kpc $< r <$ 120\,kpc with velocities $v > +275$\,km\,s$^{-1}$.
    Line-styles as in Fig.\ \ref{fig:a-imbh-inspiral}.
    $1 \sigma$ error bars of the theoretical distribution are shown in grey and are determined by bootstrapping.
    The filled bars show the distribution of all B-type HVS from the \emph{Sloan Digital Sky Survey}
    in \citet{b07a, b07b}.
    Theory would expect 2-3 HVSs with velocities $> 1000$\,km\,s$^{-1}$ which have not been found yet.
    The low-velocity bin may include stars ejected from open clusters.
  \label{fig:hvs-hist2}}
\end{figure}

We find no significant dependence of the HVS velocities produced on the IMBH mass and the density profile of the cusp.
The distribution of HVSs as they should be observable in the Galactic halo
is shown in Fig.\ \ref{fig:hvs-hist2}.
Our results show that there should be a population of HVSs with very high velocities ($v > 1000$\,km\,s$^{-1}$) which has not been observed so far.
The observed excess in low-velocity HVSs ($v<500$\,km\,s$^{-1}$) may be explained by stars ejected from young massive star clusters
as a result of binary-binary or IMBH collisions (cf.\ \citet{ggp07} and references therein).
Hence, from the small amount of data currently available, the IMBH HVS ejection model can neither be excluded nor concluded for our Galaxy.
These results are consistent with the simulations done by \citet{shm07}.

The distribution of central distances of the HVSs seems compatible with the observations.
\citet{b07a} find a slight anisotropy in the spatial distribution of HVSs;
\citet{l06} and \citet{shm06} argue that the IMBH's orbital plane gives preferred directions for HVS ejection.
However, our simulations show that the orbital
plane changes rapidly over the inspiral time of a few Myr.
Thus, the distribution of HVS directions may be fairly random, even if caused by only a few IMBH inspirals.

\subsection{HVS rotation}

HVSs are created in the Galactic Centre as a result of close encounters with massive black holes.
If these encounters are close enough, tidal effects become important.
One effect would be that, as a result of the encounter, rotation is induced in the star.

In the case of stellar binaries disrupted by the central SMBH \citep{h88},
a binary star with semimajor axis $a$ and masses $M_1, M_2$ will get disrupted at
\begin{equation}
  R_{\mathrm{tid}}^{\rm bin} = a \sqrt[3]{M_{\mathrm{SMBH}}/\left(M_1+M_2\right)} \label{eq:r-tid-bin}
\end{equation}
\citep{yt03}.
In terms of tidal radii of the member stars, $R_{\mathrm{tid}} = \sqrt[3]{M_{\mathrm{SMBH}}/M_*}\,R_*$,
this is $R_{\mathrm{tid}}^{\rm bin} / R_{\mathrm{tid}} = 0.8\,a/R_*$ for an equal-mass binary.
Thus, if a binary's semimajor axis is at least a few times the radius of each member,
a HVS created by binary disruption passes the SMBH no closer than a few tidal radii,
meaning that it would not feel a strong tidal force due to the SMBH.

In the case of HVSs created by IMBH encounters, the tidal disruption radius of a B-type star
with radius $R_* = 1.5\,R_{\odot}$ and mass $M_* = 3\,\msun$ is 
$R_{\mathrm{tid}} = 10.4\,R_{\odot}\ (15\,R_{\odot})$, assuming a IMBH mass of
$M_{\mathrm{IMBH}} = 10^3\,\msun\ (3 \times 10^3\,\msun)$.

\begin{figure}
  \begin{center}
    \includegraphics[width=8.3cm]{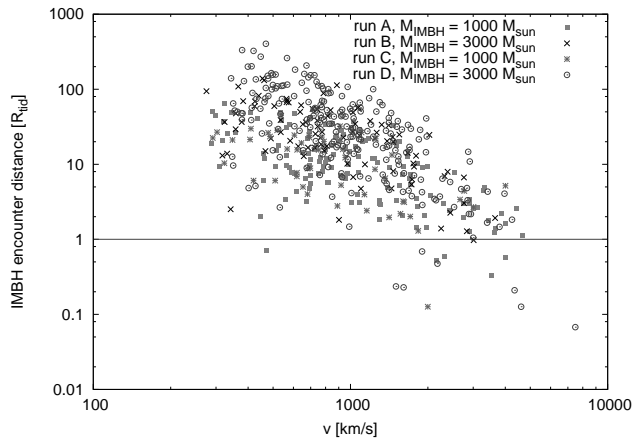}
  \end{center}
  \caption{IMBH encounter distance for ejected HVSs in units of the tidal radius,
    plotted against the HVS velocity at a distance of 10\,kpc from the Galactic Centre.
    The filled squares, crosses, asterisks, and open circles represent the data from run $A$, $B$, $C$, and $D$, respectively.
    Most HVSs created passed the IMBH at a distance of more than several tidal radii and are thus not affected by tidal effects.
    Only among the very fast moving HVSs ($v > 1000$\,km\,s$^{-1}$), a significant proportion has passed the IMBH within a few tidal radii.
    Stars passing the IMBH below $R_{\mathrm{tid}}$ will get tidally disrupted.
  \label{fig:hvs-rtid}}
\end{figure}

Fig.\ \ref{fig:hvs-rtid} shows the IMBH encounter distance of the HVSs created in our simulations.
Only among the very fast moving HVSs ($v > 1000$\,km\,s$^{-1}$),
a significant proportion has passed the IMBH within a few tidal radii.

In the following, we calculate the amount of rotation induced by a massive black hole passage, depending on the minimum distance.
Detection of rapid rotation of any observed HVS could be evidence for the presence of IMBHs in the Galactic Centre.

\begin{figure}
  \begin{center}
    \includegraphics[width=8.3cm]{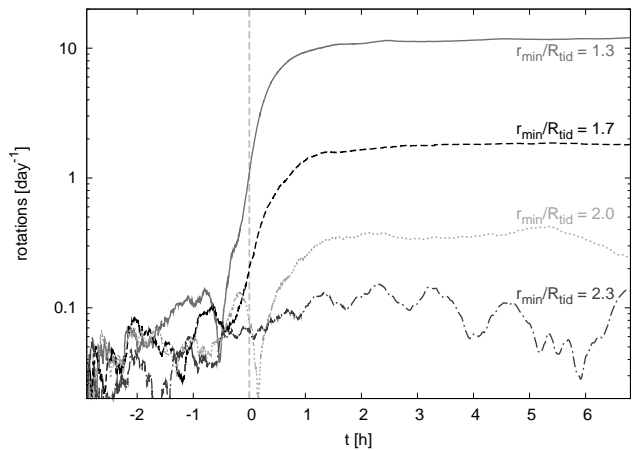}
  \end{center}
  \caption{Induced rotation of a star due to a close passage of an IMBH.
    The curves are labelled by the minimum encounter distance ($t=0$) in terms of the star's tidal radius $R_{\mathrm{tid}}$.
    Significant rotation is only induced if the minimum distance is  $r_{\rm min} \simless 2 R_{\mathrm{tid}}$.\label{fig:rot-1.5tr}}
\end{figure}

In order to test up to which encounter distance rotation is induced in a star, we made
calculations with a modified version of the SPH code of \citet{nn03}.
We simulated hyperbolic encounters of a 3\,\tmsun\ star with a radius of $R_* = 1.5\,R_{\odot}$ and a $3\times 10^3\,\msun$ IMBH.
Fig.\ \ref{fig:rot-1.5tr} shows that encounters with a minimum distance of
less than two tidal radii from the IMBH cause the passing star to rotate significantly.
An encounter within 1.3 tidal radii induces rotation with a period of two hours.

A significant rotation should be visible for those stars passing the IMBH within two tidal radii,
which in our calculations are only 3 per cent of the total number of HVSs created.
Hence, only among very fast HVSs ($v > 1000$\,km\,s$^{-1}$)
we can expect to find a significant proportion of rotating stars as evidence for an IMBH in the Galactic Centre.

\subsection{Ejection of hypervelocity binary stars}

\citet{lyl07} showed that binary stars can be ejected from the Galactic Centre as a whole through interactions with an IMBH.
Using equation (\ref{eq:r-tid-bin}) with $M_{\mathrm{IMBH}}= 3 \times 10^3\,\msun$ and  $M_1 = M_2 = 3\,\msun$,
we get $R_{\mathrm{tid}}^{\rm bin} \approx 8 a$
as the minimum distance from the IMBH which a binary star with semimajor axis $a$ could survive before being tidally disrupted.
As one can see in Fig.\ \ref{fig:hvs-rtid}, all HVSs created in our models
have encounter distances below 400 $R_{\rm tid}$, or 30\,au.
Hence, to survive these encounters, the binary stars must have a separation below 3.75\,au.

To survive the tidal forces at a distance of 0.05\,pc from a $3\times 10^6\,\msun$ SMBH,
a 6\,\tmsun\ binary must have a semimajor axis below 130\,au,
giving an upper limit for the separation of stellar binaries in the Galactic Centre.
So far, not much is known about the binary fraction $f_{\rm bin}$ and the separation distribution in the Galactic Centre.
As a rough estimate, we can use the period distribution of G-dwarf binaries in the solar neighbourhood \citep{dm91}.
We find that 37 per cent of all binaries with semimajor axis below 130\,au have a separation below 3.75\,au
and may eventually be ejected as hypervelocity binary stars (HVBSs).

Binaries ejected tend to be among the low-velocity HVSs.
If we consider the actual distribution of encounter distances as shown in Fig.\ \ref{fig:hvs-rtid},
and also the different `lifetimes' of the HVSs, i.e.\ the amount of time a star spends
in the HVS regime with velocity $v > 275$\,km\,s$^{-1}$ and distance to the Galactic Centre 10\,kpc $ < r < $ 120\,kpc,
we find that the actual fraction of HVBSs is only $0.13 \times f_{\rm bin}$, i.e.\ no more than a few per cent.

\section{Discussion}
\label{sec:discussion}

We have developed a novel post-Newtonian, high-precision integrator \textsc{bhint}
for stellar systems containing a SMBH.
The algorithm makes use of the fact that the Keplerian orbits in such a potential can be calculated directly, and are only weakly perturbed.
We used this method to follow the inspiral of IMBHs into the Galactic Centre
as a result of dynamical friction.
We find that on their way to the centre, IMBHs deplete the central cusp of stars, leaving behind a shallower density profile
compatible with the one recently observed in the centre of the Milky Way.
Due to the decreased stellar density, IMBHs may get stuck at a distance of a
milliparsec from the SMBH for a few 10\,Myr.
If the recently detected HVSs have been ejected by one or several IMBHs,
it is likely that an IMBH still resides close to Sgr A*.

Such an IMBH may be observable by detecting the proper motion of the radio source Sgr A* with upcoming
high-resolution telescopes.
Our results suggest that an IMBH in the Galactic Centre may spend several Myr in the regime where it is detectable
with an astrometric resolution of 0.1\,mas, yielding good chance of direct observation in the near future.

Once an IMBH comes close enough to the centre, gravitational wave emission will take over and lead to the merger with the central SMBH.
These mergers are among the most powerful sources of gravitational waves for the planned \emph{LISA} mission \citep{t87}
and may well be detected if the IMBH inspiral rate is of the order of 1 per 25\,Myr.

Our results show that there should be a population of HVSs with very high velocities ($v > 1000$\,km\,s$^{-1}$) which has not been observed so far,
consistent with the findings of \citet{shm07}.
However, as the sample size of HVSs observed is relatively small, IMBHs cannot be excluded as ejection mechanism.
Moreover, HVSs ejected by IMBHs are hard to distinguish from those ejected by other
mechanisms such as binary disruptions as we do not expect significant rotation in either case,
but possible observations of significant rotation of HVSs in the high velocity tail may favour the IMBH model.

\section*{Acknowledgements}
We are grateful to Douglas C.\ Heggie and Ulrich Heber for useful discussions,
and to Ingo Berentzen for his help with the implementation of the post-Newtonian extension.
We also thank Naohito Nakasato for the help with his SPH code.
This work was supported by the DFG Priority Program 1177 `Witnesses of Cosmic History: Formation and
Evolution of Black Holes, Galaxies and Their Environment'.

\makeatletter   \renewcommand{\@biblabel}[1]{[#1]}   \makeatother

\label{lastpage}

\end{document}